\documentclass{article}

\usepackage{arxiv}

\usepackage[utf8]{inputenc} % allow utf-8 input
\usepackage[T1]{fontenc}    % use 8-bit T1 fonts
\usepackage{hyperref}       % hyperlinks
\usepackage{url}            % simple URL typesetting
\usepackage{booktabs}       % professional-quality tables
\usepackage{amsfonts}       % blackboard math symbols
\usepackage{nicefrac}       % compact symbols for 1/2, etc.
\usepackage{microtype}      % microtypography
\usepackage{graphicx}
\usepackage{tgtermes}
\usepackage{eqparbox}
\usepackage[noend]{algpseudocode}

\usepackage{mathptmx}  
\usepackage[scaled=.92]{helvet}
\usepackage{amsthm,amsmath,amssymb}
\usepackage{mathrsfs}
\usepackage{todonotes}
\usepackage{algorithm,algorithmicx}
\usepackage{algcompatible}
\usepackage{libertine}

\algblockdefx[Foreach]{Foreach}{EndForeach}[1]{\textbf{foreach} #1 \textbf{do}}{\textbf{end foreach}}

\usepackage{listings}
\usepackage{enumerate}
\usepackage{enumitem}

\usepackage{wrapfig}

\usepackage[numbers]{natbib}  %% numbers is required.
\newcommand{\ra}[1]{\renewcommand{\arraystretch}{#1}}

\theoremstyle{definition}

\theoremstyle{remark}

\graphicspath{{media/}}     % organize your images and other figures under media/ folder

\title{Vulnerability Analysis of the Android Kernel
}

\author{
    Joseph R. Barr \\
    AI Group\\
    Acronis SCS\\
    Scottsdale, Arizona,
    United States of America\\
  \texttt{joe.barr@acronisscs.com} \\
   \And
    Peter Shaw\\
    Department of AI\\
    Nanjing Uni. of Information Science \& Technology\\
    JiangSu, China\\
   \texttt{peter.shaw.cs@gmail.com} \\
  \And 
    Tyler Thatcher\\
    AI Group\\
    Acronis SCS\\
    Scottsdale, Arizona,
    United States of America\\
    \texttt{tyler.thatcher@acronisscs.com} \\
}

\begin{document}
\maketitle

\begin{abstract}
We describe a workflow used to analyze the source code of the  {\sc Android OS kernel} and rate for a particular kind of bugginess that exposes a program to hacking. The workflow represents a novel approach for components' vulnerability rating. 
The approach is inspired by recent work on embedding source code functions. 
The workflow combines deep learning with heuristics and machine learning. Deep learning is used to embed function/method labels into a Euclidean space. Because the corpus of Android kernel source code is rather limited (containing approximately 2 million C/C++ functions \& Java methods), a straightforward embedding is untenable.  To overcome the challenge of the dearth of data, it's necessary to go through an intermediate step of the \textit{Byte-Pair Encoding}. 
Subsequently, we embed the tokens from which we assemble an embedding of function/method labels. Long short-term memory networks (LSTM) are used to embed tokens into vectors in $\mathbb{R}^d$ from which we form a \textit{cosine matrix} consisting of the cosine between every pair of vectors. The cosine matrix may be interpreted as a (combinatorial) `weighted' graph whose vertices represent functions/methods and `weighted' edges correspond to matrix entries.  Features that include function vectors plus those defined heuristically are used to score for risk of bugginess.
\end{abstract}


\section{Introduction}
Software is complex and invariably almost undoubtedly susceptible to exploitation by attackers who strive to penetrate computer systems, steal sensitive data, and cause significant harm. Therefore, it is important and, in some cases crucial to identify, triage, and fix vulnerabilities before attackers find and exploit them. 

The insatiable appetite for software as a vehicle for productivity \& economic growth has created a  \textit{software global software supply chain}, which by its very nature is fragmented, unregulated, with spotty standards, having a vast number of open source-software modules that are plugged into numerous commercial products, applications as well as various vital computer and electronic systems. Managing a rather haphazard supply chain system is challenging and occasionally fails to identify problems at the source code level. 

Furthermore, it's generally difficult to ascertain whether bugs are intentional. Regardless of whether deliberate or not, they create vulnerabilities that open the door for exploitation by garden-variety hackers, state-sponsored actors, and organized crime syndicates. Arguably, the most worrisome vulnerabilities are those which are continually exploited by state-sponsored actors to threaten our critical infrastructure: sectors like healthcare, energy, transportation, government, and the financial system. Accordingly, this analysis is a step developing a workflow designed to help identify vulnerabilities and, strengthen the global software supply chain.

\subsection{Terminology}
In order to simplify, we use the generic term \textit{function} to mean both function and  method (i.e., function bounded to a class.)  Sometimes we use the term \textit{component} to signify a function (or method.) We use the term \textit{deep learning}  as a stand-in for any type of neural network  architecture. In this context a \textit{classifier} is a function which estimate the conditional probability $\mathbb{P}(y=1 \ | \ x)$ for each  $x\in \mathcal{X}$ in the \textit{input set}. Sometime we add the adjective \textit{soft} and call such classifier \textit{soft classifier} to distinguish from \textit{hard classifier} which assigns a class membership to each $x \in \mathcal{X} \subset \mathbb{R}^d$.
%%%%%%%%%%%%%%%%%
%%%%%%%%%%%%%%%%%
\section{Related work}
%%%%%%%%%%%%%%%%%%%%%
Once software is deployed a production system may already be at risk. Consequently a major software supply chain issue involves vetting source code for vulnerability prior to a  deployment. With the SolarWinds attack the problem of the software supply chain was recently greatly amplified, but in fact, the problem has a long history. A well-known instance is the 2013 attack against US retail giant Target is a good example of lateral movement where hackers compromised Target’s HVAC system, then used the application’s trusted status to gain access to the retailer’s sensitive data. 

A somewhat similar technique addressing a different problem is found in a 2020 paper by Tian, et al. \cite{tian2020evaluating} where the authors compare the three types of NLP contextualized word embedding \textsc{CODE2VEC, CC2VEC} and \textsc{BERT} to differentiate between buggy code fragments and fixed and patched code. However, our aim is quite different, it is to develop tools that can identify buggy, especially malicious code in a supervised manner. For this reason we have used tagged functions from Mitre's CVE database \cite{CVE}. This CVE dataset however, has other challenges to be address such as the sparseness tagged data.

Previous work by the authors \cite{barr2020vulnerability},  \cite{barr2020combinatorial} Android's Bluetooth module Fluoride was analyzed using earlier-versioned workflow which produced results at the AUC in the 90 percent range.

\section{Methodology, a high-level view}
This empirical study of the \textit{Android kernel stack} demonstrates the feasibility of \textit{static code analysis} workflow to help identify vulnerability in source code. The workflow has several parts including function embedding, which comes in two parts: compression and deep-learning, features extraction with heuristics, and regression \& classifications. A variant of the workflow was proved effective to classify natural language \cite{barr9030927}.  Written in standard C, C++, and in Java, the Android kernel may be regarded as \textit{bags of words} consisting of tokens, one bag of tokens per function. Much like a natural language, tokens consist of names like variables, functions, literal strings, delimiters, and punctuation.  

The data, or specifically each function, is appended by a (0,1)-tag. The tags represent the presence or absence of \textit{Common Vulnerabilities \& Exposures, } CVEs \cite{CVE}. Tags are quite imbalanced; less than one percent of the records are tagged as $+1$ for CVEs.

A challenge we had encountered in learning an embedding of labels, i.e., function names, is that function names are too imbalanced to effectively learn an embedding in the Android kernel. Thus the reason for the circuitous route that goes through a `Claude Shannon-type'  \textit{data compression} phase that ultimately leads to the embedding. We've used the \textit{Byte-Pair Encoding} (BPE) \cite{Gage} algorithm to tokenize or break up functions into tokens. BPE results in much 'richer,' less imbalanced data, which can improve the capability of learning an embedding (see details below.) Although a high dimensional encoder, e.g., $d=256$ many have produced a better performance, our embedding in $\mathbb{R}^{64}$ had better performance in terms of accuracy and perplexity, for practical reasons, we have selected a vector of length $d=128$ to embed functions. In fact, we tried higher values $d$ but with $\mathbb{R}^{128}$, performance of classifier, i.e., CVE prediction, seems more than adequate. It is always important to consider the balance between bias and variance. Thus simply increasing the model complexity without noticeable improvement seems unwise. We should also mention that parsimony is always a consideration, especially for certain model frameworks some of which we have considered (e.g., Logit.) Still, we can't rule out the possibility that a higher dimensional embedding out-performs a lower dimension embedding and we indeed intend to conduct further empirical studies with higher dimension encoding in subsequent projects. The comparative performance of these embeddings are shown later in Figure \ref{fig:GBM_gain} discuss in detail later in the text after some needed discussion.

The embedding results in a `tall-and-thin' matrix with 128 columns and number of rows which equal to the number of functions. 
\begin{align}
S = \ \begin{tabular}{c}
 \itshape $f_1 \rightarrow$ \\
 \itshape $f_2 \rightarrow$ \\
 \itshape $f_3 \rightarrow$ \\
 .\\
 .\\
 .\\
 .\\
\itshape $f_N \rightarrow$ \\
\end{tabular}
\begin{bmatrix}
 s_{1,1} & s_{1,2} & . & . & . & s_{1,128} \\
 s_{2,1} & s_{2,2} & . & . & . & s_{2,128} \\
 s_{3,1} & s_{3,2} & . & . & . & s_{3,128} \\
 .      &   .   &  . & . & . & . \\
 .      &   .   &  . & . & . & . \\
 %.      &   .   &  . & . & . & . \\
 .      &   .   &  . & . & . & . \\
 .      &   .   &  . & . & . & . \\
  s_{N,1} & s_{N,2} & . & . & . & s_{N,128}
\end{bmatrix}
\end{align}

where $s_{i,j}$ is the $j$th component of the embedding of the function $f_i$ and $N$ is much larger than 128.

Heuristics-based features are extracted from the cosine matrix $S$ and the source code itself. The amalgamation of the rows of the matrix $S$ and various other features form the basis for a classifier that predict CVE tags.  
The artifacts of the workflow are:
\begin{enumerate}
\setlist[enumerate]{itemsep=1mm}
    \item Data compression with Byte-Pair Encoding results in a bag of tokens; one bag of tokens for each function.
    \item An embedding, or \textit{auto-encoding} with LSTM; a vector in $\mathbb{R}^{128}$ for each function.
    \item Heuristics-based feature a vector, one for each function.
    \item Risk rating, a collection $(f, p_f)$, where $f$ is a function of Android and $p_f$ represents the level of risk, or exposure associated with the function $f$.
\end{enumerate}

\begin{figure}[t!]
   \centering
   \includegraphics[width = 5.5 cm]{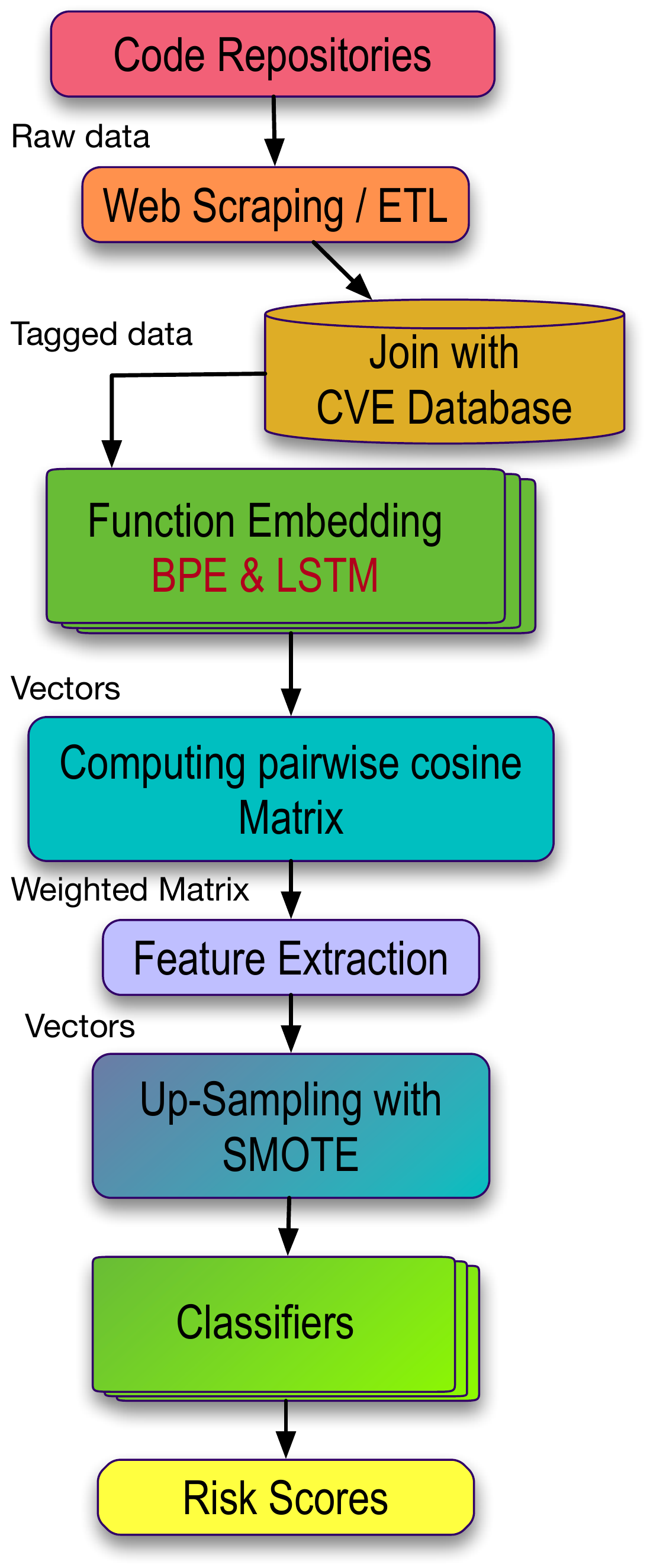}
  \caption{Process tool chain.}
   \label{fig:toolchain}
%\vskip -10pt   
\end{figure}

\section{The process workflow}

Figure \ref{fig:toolchain} describes the toolchain used to classify the code  extracted from the code repositories. This tool chain modifies one previously developed by the authors \cite{barr9030927}. The more recent one being \cite{barrTransAI2020}. 

The toolchain begins with input from the code repository, which is the entry point shown in Figure \ref{fig:toolchain}. This raw code is merged with the CVE database and then tokenization using our Byte-Pair Encoding algorithm to produce new tokens followed by \textsc{embedding} which embed [sic] `sub-tokens' using the LSTM network, which is then used to `assemble' our function embedding.  You can find LSTM tools at \cite{LSTMSciTools2}.

The embedding consists of 128-dimensional vectors, one for each function. Thus, embedding results in a matrix $A$, say, whose rows represent the functions and columns are the embedding `directions'. The matrix is `very tall and skinny,' having as many rows as there are functions, and by design, it has 128 columns.  We've experimented with various embedding and opted for 128 dimensions for practical reasons. 

Having produced the matrix $A$ we then compute a pairwise-similarity matrix $S$ where the $(i,j)$-entry $s_{ij}$ is \vskip -10pt
\begin{align}
\cos (a_i, a_j)&=\dfrac{a_i^Ta_j}{\|a_i\|\|a_j\|}
\end{align}
where $a_k$ is the $k$th row of $A$. 
Since the number of functions is 1,154,416, the matrix $S$ is of order 1,154,416-by-1,154,416.

Recall that every row of $A$ is a function embedding. Thus, we may think of the input space $\mathcal{X}$ as the space of all 1,154,416 $R^{128}$-vectors. 

Heuristically-inspired features are extracted, which result in a feature vector, one for each function.  A list of features is:
\begin{itemize}
    \item Function length; the number of tokens per function.
    \item The length of longest line in a function. (Since we're in the C/C++ domain a line is well defined.)
    \item The density of tokens shared with the aggregates of CVEs.
    \item Row sums of the cosine matrix. (This feature is a stand-in for the degree of the vertex.)
    \item The number of parameters of the function.
\end{itemize}
Append those five heuristically-derived features to the $\mathbb{R}^{128}$ auto-encoded features. This nets a 133 feature vector which forms a basis for the classifier.

The final phase is modeling for vulnerabilities, i.e., calculating the conditional probability $\phi(x)=\mathbb{P}(y=1 \ \| \ x)$, that a function is vulnerable given feature vector $x$. We estimate $\phi(x)$ from training pairs $(x,y)$ where $x\in \mathbb{R}^{133}$ and $y=0$ if the function is not CVE and $y=1$ otherwise. There are 133 features, which include the 128 auto-encoded features plus, five which were produced heuristically. 

Since CVE tags are imbalanced, we employ \textit{SMOTE}, an up-sampling technique \cite{2002Smote}.   SMOTE has a parameter that controls the percentage of synthetic positive examples. We tried various values, and a 20\% synthetic positive example is optimal. 

The outputs of the workflow are:
\begin{enumerate}
\setlist[enumerate]{itemsep=2mm}
    \item vector representation of functions,
    \item features vector, and
    \item vulnerability rating.
\end{enumerate}

\section{The data}
This analysis is an empirical study using the Android OS Stack v.10, which is written in C, C++, and Java. Android is an open-source project; download instructions are found in \cite{Android1} and with Android components in \cite{Android2}.

The code is spread over 103,111 *.c, *.cc, *.cpp and *.h files, and with Java consists of 83,437 files. The total number of functions is 1,154,416. The number of *.java files is 83,437 and the number of functions (methods) is 1,266,073.

There are 6,775 CVE associated with the C and C++ code, while only 680 CVEs are associated with the Java code. 

A summary in tables below in table \ref{tab:codesample} and a link to the source code in the reference \cite{Android}.

%\tabcolsep}{2pt}
\begin{table}[h]
\begin{center}
%\captionsetup{justification=centering}
\setlength{\tabcolsep}{4pt}
\caption{\textsc{Representative languages sizes in the sample code}\label{tab:codesample}}
\vspace{3pt}
\ra{1.2} 
\begin{tabular}{c|r|r|r|r}
    \toprule
    & \textbf{.c} & \textbf{.cpp} & \textbf{.cc} & \textbf{Java}\\
    \midrule
    No. of Files & 36,887 & 51,172 & 15,052 & 83,437 \\
    Total Size (approx. MB) & 595 & 974 & 295 & 891\\
    \bottomrule
\end{tabular}    
\end{center}
%\vskip -10pt
\end{table}

\textsc{An itemized list of the code composition is provided in Table \ref{tab:composition} below.}

\begin{table}[h!]

\begin{center}
\caption{\textsc{Sample code statistics}
\label{tab:composition}}
\vspace{3pt}
\ra{1.2}
\setlength{\tabcolsep}{8pt}
    \begin{tabular}{l|r}
    \toprule
    \textbf{C \& C++ code}   & \textbf{Totals} \\
        \midrule
    Lines of code  & 39,353,232  \\
    Number of files & 103,111   \\
    Size (approx. in MB)  & 1,864 \\
    Number of functions   & 1,154,416 \\
    Tokens count     & 204,777,237 \\
    Unique tokens count  &  3,964,717 \\
    Number of CVEs  &  6,775\\
    \bottomrule
    \end{tabular}
%\vskip -10pt
\end{center}

\begin{center}
\setlength{\tabcolsep}{8pt}
\begin{tabular}{c|r }
\toprule
   \textbf{Java} & \textbf{Totals}   \\
    \midrule
    No. of Files & 83,437  \\
    Total Size (in MB) & 891  \\
    Number of functions (methods) & 1,266,073 \\
    Number of CVEs  & 680\\
\bottomrule
\end{tabular}    
\end{center}
%\vskip -15pt
\end{table}

%%%%%%%%%%%%%%%
\section{Common vulnerability exposures}
Common Vulnerabilities \& Exposures (CVE) is a database of publicly-disclosed software \& firmware vulnerabilities, i.e., risky for exposure to malware \cite{CVE}. To illustrate, below are a few examples of functions tagged as CVEs. 
\noindent
\textsc{Example 1.} \texttt{CVE-2017-0781} Code inspection of \cite{CVE-2017-0781}.

\begin{lstlisting}[language=C,basicstyle=\scriptsize]
void bnepu_release_bcb(tBNEP_CONN* p_bcb) 
{   
  /* Ensure timer is stopped */
  alarm_free(p_bcb->conn_timer);
    ....
    ....
  while(!fixed_queue_is_empty(p_bcb->xmit_q)) 
  {
    osi_free(fixed_queue_try_dequeue(
    p_bcb->xmit_q));
  }
  fixed_queue_free(p_bcb->xmit_q, NULL);
  p_bcb->xmit_q = NULL;
}
\end{lstlisting}
\normalsize 
\begin{figure}[!ht]
      \centering
      \vskip -20pt
      \includegraphics[width = 8 cm]{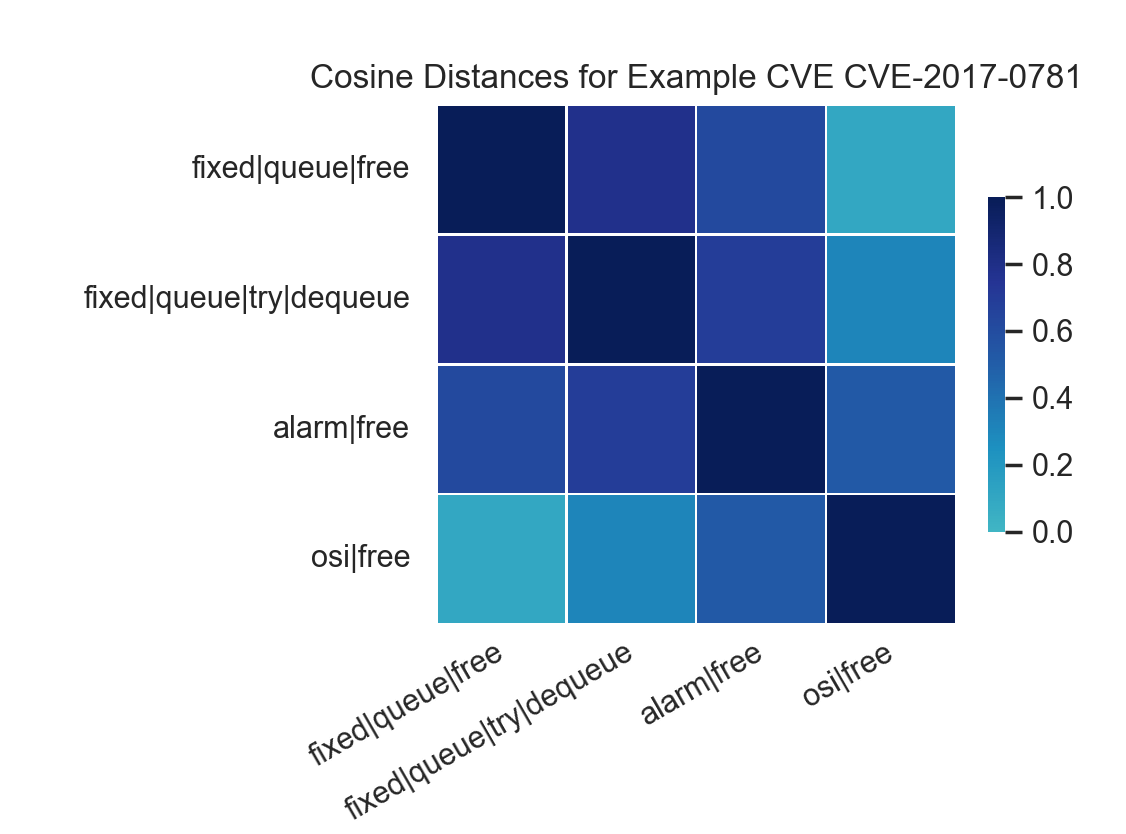}
      \caption{The pairwise similarity between function present in  CVE-2017-0781.}
      \label{fig:cve_correlations}
      %\vskip -15pt
\end{figure}

A plausible argument why CVE-2017-0781 might pose a vulnerability is that the \texttt{free} commands have different names, but seem to free the same memory block. The similarity was evident in the $\cos$ distances between the corresponding vectors in our embedding. These similarities are shown in Figure \ref{fig:cve_correlations}.

Based on the `commit' message which says 
\texttt{Free p\_pending\_data from tBNEP\_CONN} to  
\texttt{avoid potential memory leaks}, i.e., CVE-2017-0781 cause was a memory leak.  \par 
%\vskip 3pt
\noindent\textsc{Example 2.} \texttt{CVE-2020-0240} \cite{CVE1}\par
For CVE-2020-0240, the `commit' message {\scriptsize \texttt{``Fix integer overflow in NewFixedDoubleArray''}} 
\normalsize
says it all.

\begin{lstlisting}[language=C,basicstyle=\scriptsize]
Handle<FixedArrayBase> Factory::NewFixedDoubleArray(
                int length,
                    PretenureFlag pretenure) {
  DCHECK_LE(0, length);
  if (length == 0) return empty_fixed_array();
  // Should be if (length < 0 || length > 
  //    FixedDoubleArray::kMaxLength)
  if (length > FixedDoubleArray::kMaxLength) {
    isolate()->heap()->FatalProcessOutOfMemory(
           "invalid array length");
  }
  int size = FixedDoubleArray::SizeFor(length);
  Map* map = *fixed_double_array_map();
  HeapObject* result =
      AllocateRawWithImmortalMap(size, pretenure, map, 
            kDoubleAligned);
  Handle<FixedDoubleArray> array(FixedDoubleArray::cast(
            result), isolate());
  array->set_length(length);
  return array;
}
\end{lstlisting}
%\vskip 5pt
\noindent\textsc{Example 3.} \texttt{ CVE-2019-20636} \cite{CVE2}
\\
For CVE-2019-20636 the `commit' message is \par
{\scriptsize
\noindent\texttt{``If we happen to have a garbage in input  device's}\par
\noindent\texttt{keycode table with values too big we'll end up doing}\par \noindent\texttt{clear\_bit()} \texttt{with offset way outside of our}\par \noindent\texttt{bitmaps,  damaging other objects within an input device}\par
\noindent\texttt{or even outside of it. Let's add sanity checks to the}\par \noindent\texttt{returned old  keycodes.''}
}\par

\begin{lstlisting}[language=C,basicstyle=\scriptsize]
static int input_default_setkeycode(
          struct input_dev *dev,
		  const struct input_keymap_entry *ke,
		  unsigned int *old_keycode)
{
	unsigned int index;
	int error;
	int i;
	if (!dev->keycodesize)
		return-EINVAL;
	if (ke->flags & INPUT_KEYMAP_BY_INDEX) {
		index = ke->index;
	} else {
		error = input_scancode_to_scalar(ke,&index);
		if (error)
		  return error;
	}
	if (index >= dev->keycodemax)
		  return -EINVAL;
	if (dev->keycodesize < sizeof(ke->keycode) &&
		  (ke->keycode >> (dev->keycodesize * 8)))
		   return -EINVAL;
	switch (dev->keycodesize) {
		case 1: {
		  u8 *k = (u8 *)dev->keycode;
		  *old_keycode = k[index];
		  k[index] = ke->keycode;
		  break;
		}
		case 2: {
		  u16 *k = (u16 *)dev->keycode;
		  *old_keycode = k[index];
		  k[index] = ke->keycode;
		  break;
		}
		default: {
		  u32 *k = (u32 *)dev->keycode;
		  *old_keycode = k[index];
		  k[index] = ke->keycode;
		  break;
		}
	}
    \\clear_bit() may have out-of-bounds memory access 
    \\if the keycode value is too big. Wrap clear_bit()
    \\in a sanity check: if (*old\_keycode <= KEY_MAX)
   __clear_bit(*old_keycode, dev->keybit);
   __set_bit(ke->keycode, dev->keybit);
   for (i = 0; i < dev->keycodemax; i++) {
    if (input_fetch_keycode(dev, i) == *old_keycode) {
	  __set_bit(*old_keycode, dev->keybit);
   break; /*Setting the bit twice is useless, so break*/
   }
  }
  return 0;
  }
}
\end{lstlisting}

\noindent\textsc{Example 4}. \texttt{CVE-2020-0103}\cite{CVE3}\\
For CVE-2020-0103 the `commit' message is 
{\scriptsize
\texttt{`Memory allocated by osi\_malloc() should be freed by osi\_free().'}
}\normalsize
\begin{lstlisting}[language=C,basicstyle=\scriptsize]
void a2dp_aac_decoder_cleanup(void) {
  if (a2dp_aac_decoder_cb.has_aac_handle)
    aacDecoder_Close(a2dp_aac_decoder_cb.aac_handle);
  free(a2dp_aac_decoder_cb.decode_buf);
  memset(&a2dp_aac_decoder_cb, 0, 
          sizeof(a2dp_aac_decoder_cb));
}
\end{lstlisting}

Our goal is to classify CVE tags; however, since CVEs are incredibly imbalanced, the challenge of finding the proverbial needles in a haystack is significant.

\section{Embedding}
\textit{Word embedding} forms the basis for natural language processing. Sometimes the term \textit{auto-encoding} is used to describe embedding, so an algorithm associated with embedding is referred to as an \textit{auto-encoder}. Embedding a vocabulary means to map words or tokens into a Euclidean space of dimension much smaller than the size of the vocabulary: $w \rightarrow x=(x^w_1, x^w_2,..., x^w_k)^T$ where $w$ is a word in the vocabulary and $x^w_j \in \mathbb{R}$. The mapping preserves relation between words; vectors representing words having syntactic relation tend to be `close,' and conversely, vectors of words with little or no syntactic relation do not. In general, word embedding employs a supervised algorithm that involves constructing of a neural network of one kind or another \cite{mikolov2013distributed}.

\subsection{Background}
Word embedding with \textit{word2vec} \cite{mikolov2013efficient} is a  standard technique in text analysis.  Word2vec is an implementation of neural networks to represent words as \textit{dense} vectors. The adjective `dense' refers to the fact that the vector's dimensionality is an order of magnitude smaller than the size of the dictionary. With \textit{negative sampling}, text may be trained efficiently, and when coupled with an appropriate prepossessing and  standardization, embedding with word2vec effectively represent relations between words.  Since its inception, word2vec has  brought about a myriad of applications of word embedding to practical problems, from application to indexing \& searching in a corpus of documents, to document classification, and even to real estate valuation e.g., \cite{ShahbaziBHS16}, \cite{barrTransAI2020}.

Neural networks or `deep learning' are standard tools for dealing with natural languages. The idea goes back to work by Bengio et al. \cite{bengio2003}, and later Mikolov et al.  \cite{mikolov2013distributed}. Subsequently, deep learning was applied to static code analysis in 2019 by Alon et al., \cite{alon2019code2vec} who used neural network techniques to embed code into a Euclidean space of an appropriate dimension. 

\subsection{Function Embedding}
\begin{figure}[h!] 
  \centering                            
  \includegraphics[scale=0.5]{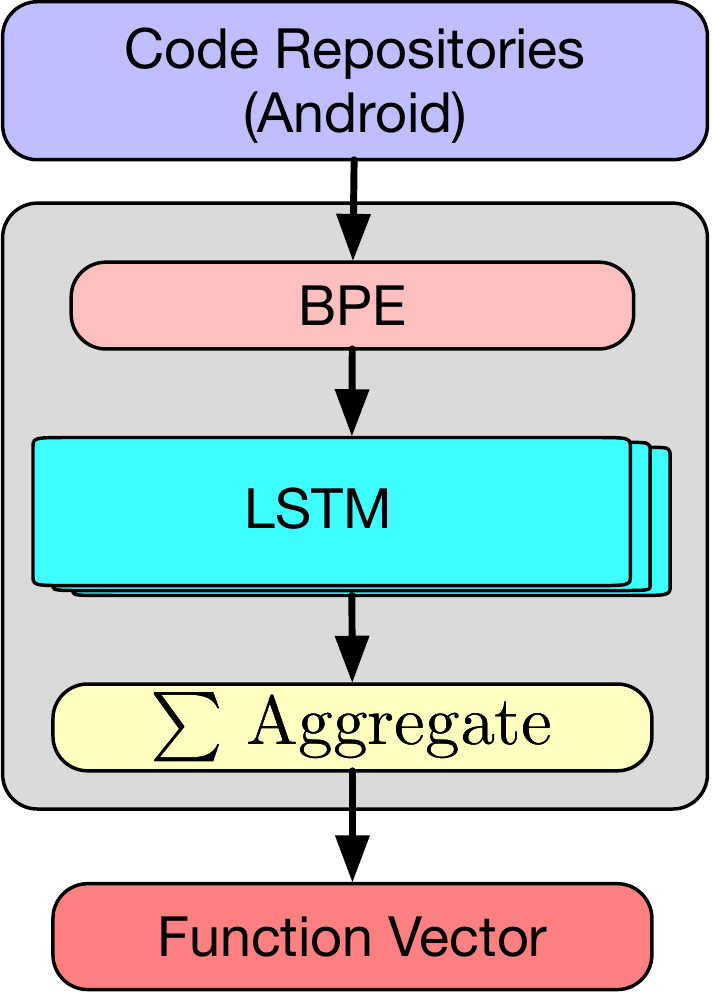} 
  \caption{Function embedding.}
  \label{fig:figureA}
  \vskip -7pt
\end{figure} 
This section introduces the idea of generating function embedding. A  \textit{function embedding} is a `summary' of a function's body; it's a fixed-length vector $x\in \mathbb{R}^k$ preserving the syntactic and semantic meaning of a function. Alternatively, function embedding is a mapping $f\rightarrow x_f\in \mathbb{R}^k$. We treat programming languages (Java, C, C++, Python, etc.) as natural languages; therefore, using a natural language model (NLM) is apt. To generate a function embedding, we first tokenize the source code of a function via \textit{ Byte Pair Encoding} (BPE), a data compression algorithm (See Figure \ref{fig:figureA}.) With \textit{long short-term memory} networks (LSTM), tokens are then transformed into `dense' vectors, which in turn are rolled up to form function embedding.
While numerous LSTM models are now available. For this work we used the standard LSTM model (as defined by \cite{jozefowicz2015empirical} section 2)
Schematically,

\subsection{Byte-Pair Encoding}
In natural language processing (NLP) models, the inputs are sequences of sentences where each sentence consists of whitespace-delimited words or tokens. Whitespace tokenizer works well for language models because the number of distinct words in a vocabulary is rather limited, or fixed, at the least. However, when it comes to computer languages, in source code processing, the vocabulary isn't constrained; there are potentially infinitely many unique variable names and strings. This leads to the  \textit{out-of-vocabulary} (OOV) problem where rare tokens (those occurring once or twice in the entire corpus) aren't learnable.

Experimentation shows that even with a rather small vocabulary size, e.g., a vocabulary of 10,000, about half of the tokens are out of the vocabulary. After comparing multiple tokenization approaches, Karampatsis et al. \cite{Karampatsis}  showed that \textit{Byte-Pair Encoding} (BPE) \cite{Gage} results in good performance across various evaluations. This is the main reason why in this project, we have used BPE to tokenize source code. 

Originally Gage \cite{Gage} applied a BPE algorithm for data compression. We use a slightly modified Byte-Pair Encoding algorithm \cite{Sennrich} to merge frequent sub-word pairs instead of replacing those with another byte. At a high-level, each word is split into characters; then, the most frequent consecutive characters pair up into a single token. This merge process repeats until it reaches the desired number of merge operations or until the desired vocabulary size is achieved. Details and examples are found in \cite{Sennrich}. Consequently, a variable name like ``dwReadSize'' is likely split into tokens [``dw'', ``Read'', ``Size'']. While ``dw'' may or may not in the vocabulary, the rest are almost certainly in the vocabulary. Thus the procedure preserves the `meaning' of that particular variable name.

We've noticed that for 897 functions, the tokenizer failed to yield consistent results. It turned out that the reason for that is that a comma occurs naturally in a    `template-scoping' operator. (Templates are C++ constructs.) The following example demonstrates the phenomenon\\ 
\noindent{\scriptsize \texttt{IntraPredBppFuncs\_C<block\_width, \\ block\_height, bitdepth, Pixel>::DcFill}}.    
\normalsize

This template form isn't an anomalous code. Indeed, the aforementioned template function takes three parameters, and the scope of the function \texttt{DcFill}  dictates its behaves precisely as specified in the class template. It's amply clear that this is a feature of C++, not an anomaly or an unusual coding standards.  To recover from ``embedded'' commas, we've used a tab (ASCII DEC-9) as a delimiter, which reduced `misfires' to a  manageable number of approximately 30.

As is the case with any natural or human language, anomalies are inevitable. Still, it seems that in our case, using a tab as a delimiter (in function names, exclusively) is as optimal as one would reasonably tolerate.

\subsection{Long Short-Term Memory (LSTM) Networks}
Once the Byte-Code Encoding is implemented on the corpus, each token is represented by a unique integer, a \textit{token id}. This integer encoding represents the natural ordering of words, which, in our case, is not ideal. Instead, we use a \textit{learnable embedding layer} to convert token IDs to `dense' vectors.

\begin{figure*}[!ht]
   \centering
   \includegraphics[width = 13.0 cm]{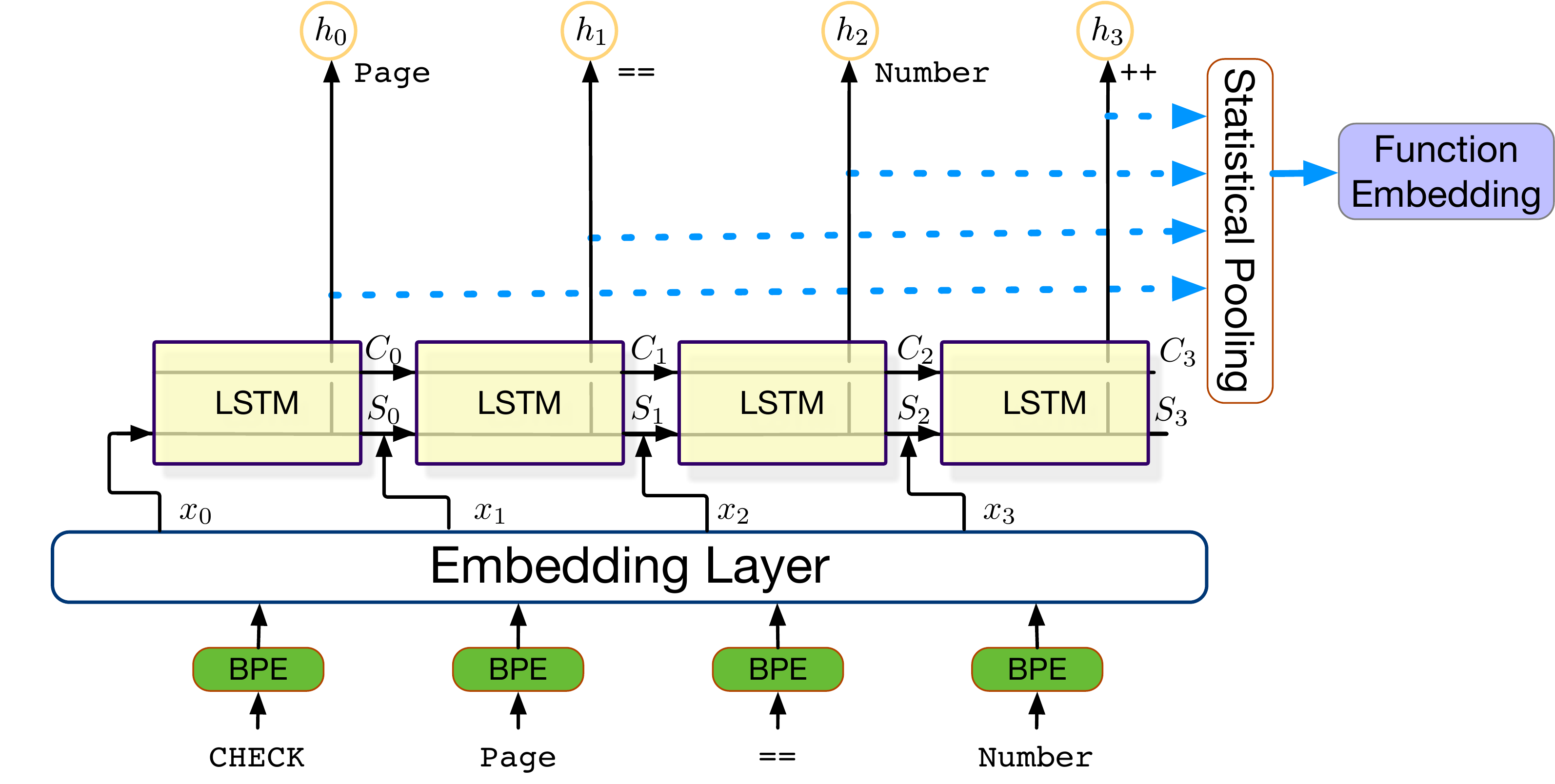}
   \caption{LSTM language model for generating function embedding}
   \label{fig:LSTM}
   \vskip -7pt
\end{figure*}

As Figure \ref{fig:LSTM} demonstrates, the dense vectors are fed to an LSTM layer \cite{Hochreiter}, which generates a hidden state $h$ for each token $t$. Roughly, if $t$ is the `pivot' token, then the learning process takes into account all the token information corresponding to $t$. Finally, averaging is used to aggregate all the hidden states to generate the function embedding.

Training this kind of LSTM network is not dissimilar to other language models. In virtually all language models involving a neural network, a (properly encoded, e.g., `one-hot') sequence of tokens is fed to the language model, where the \textit{target output} of each input token is the next token of that input. 

In our case, given a token sequence $w$, at time step $i$, the model calculates a hidden state $h_i$ based on the word embedding $x_i$ of $w_i$ and the previous hidden state $h_{i-1}$. 

The `hidden state' goes through a fully-connected layer or \textit{decoder} and is activated by the SOFTMAX function to produce a probability vector of a token to be predicted.  We proceed recursively and use the `true' next token information to train the model parameters, which maximize the accuracy of the following token predictions. \textit{Cross-entropy} is the standard cost function. 

\subsection{A brief note on code2vec}
Although we didn't use the \textit{code2vec} embedding technique, we find it worthy of mention because it is state-of-the-art technology relating to function embedding. Since code2vec requires large corpora consisting of multiple projects (See \cite{alon2019code2vec}) and since ours is a corpus consisting of a single project, the \textit{code2vec} technique didn't quite fit our needs.

Developed in 2018 by Alon et al., the \textsc{Code2vec} model framework assigns vectors to functions aiming to capture subtle differences between syntactically similar functions \cite{alon2018codeseq}. The model first parses a function and constructs an \textit{abstract syntax tree (AST)}. Then, the AST is traversed, and paths between terminal nodes, or leaves, are extracted. Each path and its corresponding terminal nodes is mapped to an embedding, a vector in $\mathbb{R}^d$. The model also has a \textit{path-attention network} to assign a score to each embedding and aggregate embedded vectors into a function vector using the \textit{attention score}. During the evaluation phase, the model computes similarities between the function vector and every function name embedding and predicts a possible function name. 

\section{Calibration of hyper-parameters}
In the following subsections, we describe how to optimized the \textit{hyperparameters} of the LSTM network. 
%%%%%%%%%%%%%%%%

For the LSTM model, we first bootstrapped a training set and formed a `set difference' of the dataset, which resulted in training and test datasets. We've tried several combinations of token embedding sizes and
evaluated the token prediction accuracy and \textit{perplexity}. 

\begin{figure}[ht!]
   \centering
   \includegraphics[width = 9.0 cm]{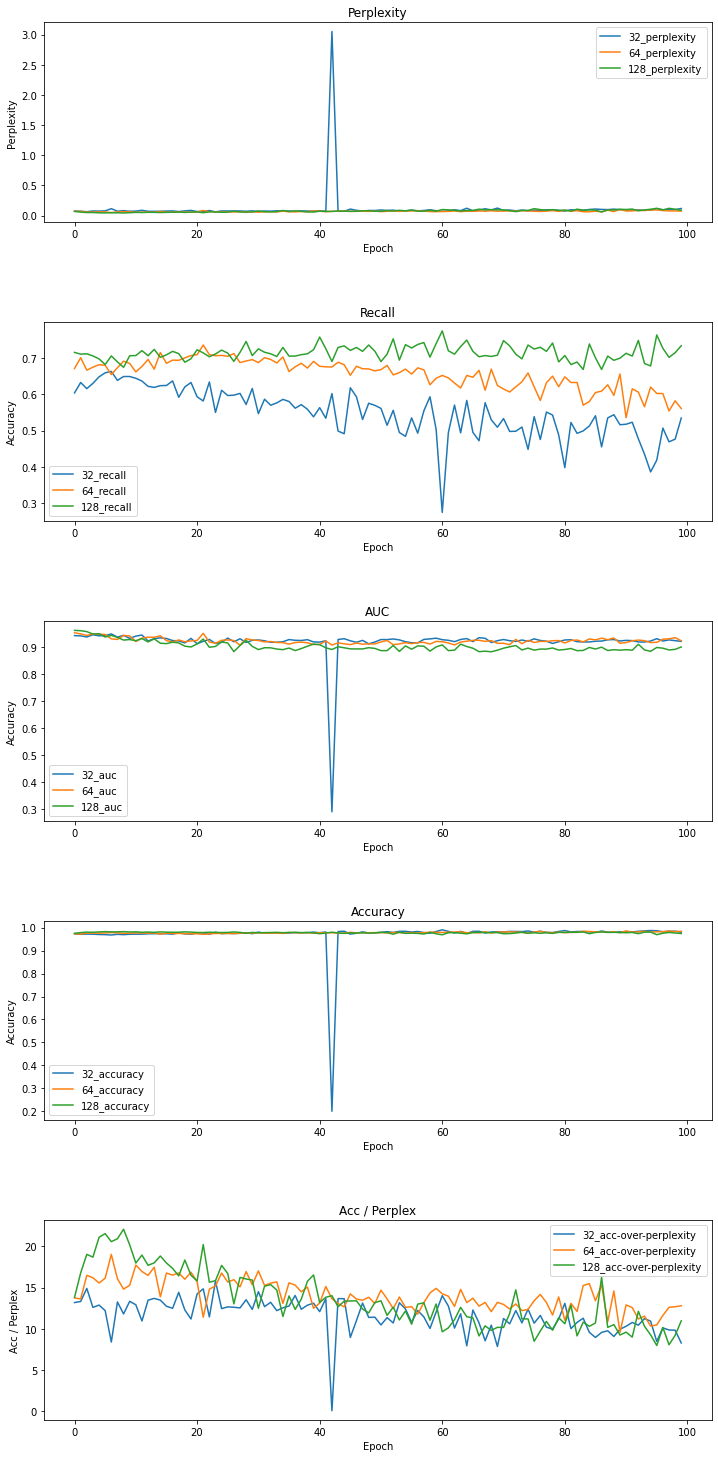}   
  \caption{Comparison of Embedding Metrics for $\mathbb{R}^{32}$, $\mathbb{R}^{64}$ and $\mathbb{R}^{128}$.}
   \label{fig:NONCVEAccuracy}
\vskip -20pt   
\end{figure}

As Figure \ref{fig:NONCVEAccuracy} shows, if the size of the token embedding and the output vector was too large, the perplexity on the test set first decreased and later increased, indicating an overfit. We've tested embedding between $\mathbb{R}^{128}$ and embedding into $\mathbb{R}^{128}$. The latter ($\mathbb{R}^{128}$) seems appropriate to `splitting the difference' on accuracy and perplexity.

\section{Classification}
As was mentioned in the `terminology' section, a  classifier is a mapping $\Phi: \mathcal{X} \rightarrow [0,1]$ that estimates the conditional probability $\phi(x)=\mathbb{P}(y=+1 \| x)$ which is often referred to as a \textit{score}, or \textit{the score of x}. We'll stick with that terminology.

As we noted earlier, the Android data is extremely imbalanced; of the 2,420,489 functions (C, C++, and Java functions), 7,455 approximately 0.3\% are positively tagged while the remaining 99.7\% are negatively tagged.  For practical reasons we're focused attention on the 1,154,416 C/C++ functions, of, which 6,775 are tagged as CVEs, i.e only approximately 0.6 of a percent of the lines were tagged. 

The goal is to predict CVEs from feature vector $x$, i.e., to estimate $\mathbb{P}(y=1 \ | \ x)$ where $$y = \begin{cases} 1 & \text{ if } y \text{ is CVE } \\
0 & \text{ else }
\end{cases}$$ 

In order to train a classifier which is based extreme imbalance we employ an \textit{up-sampling} technique  \textit{SMOTE: Synthetic Minority Over-sampling Technique} \cite{2002Smote}. SMOTE  simulates  positively-labeled tags based on similarity  to positively tagged examples.

\subsection{Baseline classifier}
\begin{figure}[t!]
   \centering
   \includegraphics[width = 8.0 cm]{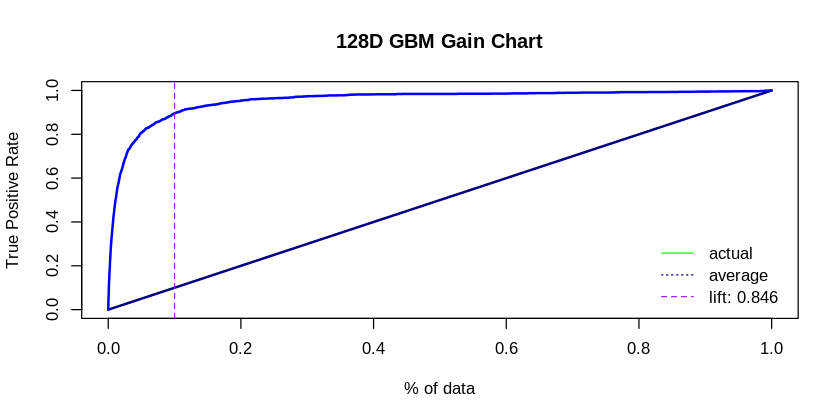}
  \caption{Baseline classifier: using $\mathbb{R}^{128}$ embedding with GBM.
   \label{fig:Gain128}}
\vskip -10pt   
\end{figure}
To establish a rough baseline, we use function embedding, the $\mathbb{R}^{128}$ function-vectors as a basis for a classifier. We use GBM as a baseline with performance in the table \ref{tab:gbm} below.
Also, see %figure \ref{ref:gbm} and
Figure \ref{fig:Gain128}.

\begin{figure*}[hb!]
   \centering
   \includegraphics[width = 12.0 cm]{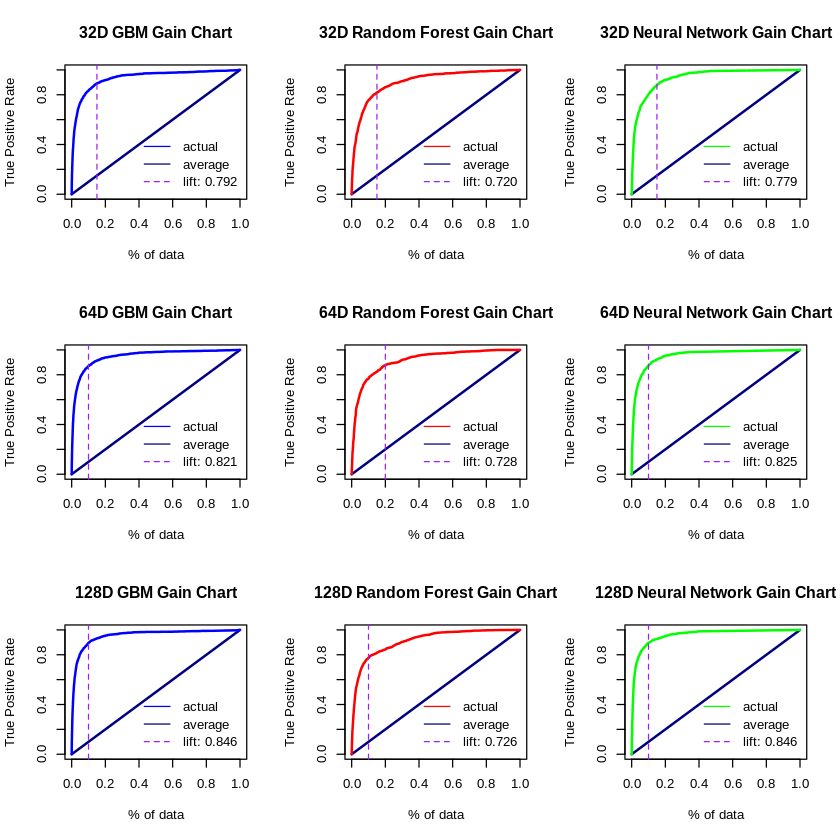}   
  \caption{Gains and lift across different embedding dimensions.}
   \label{fig:GBM_gain}
\vskip -10pt   
\end{figure*}

\begin{table}[ht]
\setlength{\tabcolsep}{3pt}
\ra{1.2}
\begin{center}
\caption{\textsc{Baseline classifier performance} \label{tab:gbm}}
\vspace{3pt}
    \begin{tabular}{cr}
    \toprule
    \textbf{Metric}   & \textbf{Performance} \\
        \midrule
      AUC & 0.813 \\
    Precision & 0.243 \\
    Specifity & 0.986 \\
     Lift & 0.868 \\\bottomrule
    \end{tabular}
    \end{center}
    \vskip -10pt
\end{table}

\subsection{Heuristics-based features}
We introduce six additional features.

\begin{enumerate}
\setlist[enumerate]{itemsep=1mm}
\item \textit{Function length}, i.e., number of tokens. 
\item \textit{Token prevalence}, i.e., percentage of `bad' tokens in a function. 
\item \textit{Row sums} of cosine matrix. 
\item \textit{Length of the longest line}, i.e., number of tokens. 
\item \textit{Number of function parameters} (parameters passed to function.)
\end{enumerate}

The first feature, \textit{Function length}, list item 1 above, aligns with the `long-method' code-smell of Kent Beck and Martin Fowler \cite{fowler1999refactoring, fowler2018refactoring}. The value used for this feature is summarized in the {\sc Function Length} Algorithm \ref{alg:alg1}. The \textit{number of tokens} used refers to the number of `bag' tokens, i.e., the size of a multi-sets. Every token is counted as many times as it appears in the function. If, for example, a token \texttt{myNum} occurs seven times, it increments the count seven times.

\begin{algorithm}[hb]
%\SetAlgoLined
 \caption{{\sc Function Length}\label{alg:alg1}}
 \hspace*{\algorithmicindent} \textbf{INPUT:} {A corpus $\mathcal{C}$ of functions}
\begin{algorithmic}[1]
    \Foreach{function $f \in  \mathcal{C}$} 
        \State Let $F(f) \leftarrow$  bag of tokens in $f$ 
        \State Let $fSize \leftarrow 0 $
        \Foreach{$t \in F(f)$} 
           \State {$fSize\leftarrow fSize +1$} 
        \EndForeach
        \State Return $fSize$ 
    \EndForeach
\end{algorithmic}
\end{algorithm}
%\vskip -5pt
The second feature listed as item 2, \textit{Token prevalence} is the most frequent `meaningful' token occurring in CVE functions, i.e., in the sense of the tokens which are common to CVEs, but less common in non-CVEs. The algorithm used to obtain the token prevalence is presented in Algorithm \ref{alg:alg2}.
The fifth feature, the \textit{Number of function parameters}, is indicative of the code-smell ``Too many parameters'' is a nominal value; say the function \texttt{myFunc}, \texttt{void myFunc(int, int, double, char*)}
has four (4) parameters, although \texttt{char*} could very well pass a long array of type \texttt{char} characters.

\begin{algorithm}[h!]
%\SetAlgoLined
 \caption{{\sc Token Prevalence}\label{alg:alg2}}
 \hspace*{\algorithmicindent} \textbf{INPUT:} {A corpus $\mathcal{C}$ of functions, and}\\
 \hspace*{\algorithmicindent} %\textbf{INPUT:}
 \:\:\:\:\:\:\:\:\:\:\:\:{A trimmed lexicon $\mathcal{L}$ of CVE Tokens}
\begin{algorithmic}[1] 
 \Foreach{function $f \in \mathcal{C}$}%{\
    \State $F(f) \leftarrow$  bag of tokens in $f$
    \State $N(F(f)) \leftarrow$ token count of $F(f)$
    \State $k \leftarrow 0$
    \Foreach{$t \in L$}
        \If{token $t$ appears  in $F(f)$} 
        \State {$k\leftarrow k+1$}
        \State Return $\Big(k/N(F(f))\Big)$ 
    \EndForeach
 \EndForeach 
\end{algorithmic} 
For each function $f$, Algorithm \ref{alg:alg2} returns the frequency of tokens in $L$ lying in the bag of tokens $F(f)$.
%\vskip -25pt
\end{algorithm}

\subsection{A trimmed CVE lexicon}
To amplify the generation of feature 2, we construct a table of all CVE tokens, i.e., a bag of tokens appearing in functions labeled as CVEs, and their frequency. Clearly, tokens like delimiters and punctuation invariably are the most frequent, and arguably those provide little or no information about the nature of the function. The lexicon is constructed by trimming the most and the least frequent tokens. Those tokens in the `middle' consist of trimmed CVE lexicon $\mathcal{L}$. The size of the intersection of the lexicon with a function is feature 2), `token prevalence'.   

\section{CVE Classification}
As noted, the transformed data consists of approximately 1.1 million $(X,y)$-pairs with $X\in \mathbb{R}^{133}$ and $y\in \{+1, -1\}$. The vector $X=(x_1,...,x_{128},x_{129},...,x_{133})$ with $x_1,...,x_{128}$ are auto-encoded and the last five $x_{129},...,x_{133}$ are heuristics-based features. Although one could try other model frameworks like support-vector machines, for practical reasons gradient boosting (GBM), random forests and neural networks were at play.
To ensure consistency of results two approaches were investigated.

\begin{enumerate}
\setlist[enumerate]{itemsep=1mm}
    \item {\sc SMOTE} which was optimized the size of up-sampled positive examples, a 20 percent up-sample rate was found optimal for the given data.
    \item {\sc The bootstrap} which re-samples data with replacement, was also deployed. The training set for {\sc The bootstrap} consisted of all the positive labels together with an equal number of negative labels sampled uniformly with replacements. We iterated the process five times. In fact, this variant of the bootstrap was done solely as a `sanity check' to ensure that results didn't significantly deviate from what was expected. However, SMOTE had consistently outperformed the bootstrap and consequently we abandoned the methodology in favor of SMOTE. 
\end{enumerate}

\subsection{CVE classifier performance summary}
\begin{figure*}[ht!]
   \centering
   \includegraphics[width = 14.5 cm]{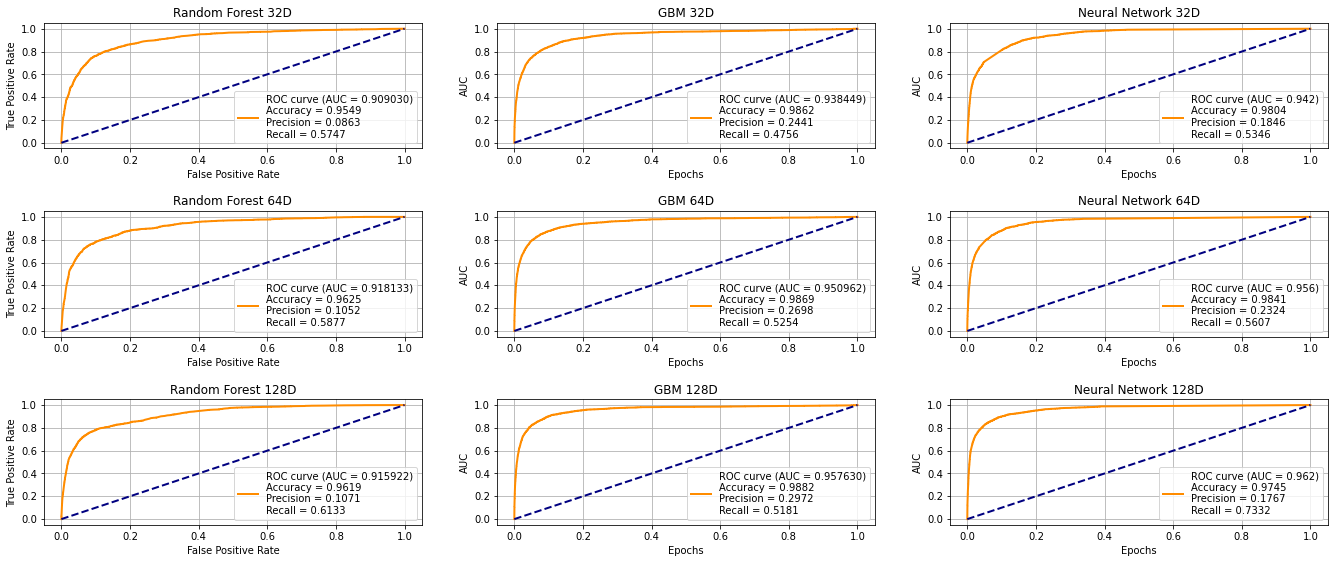}
  \caption{Prediction across multiple embedding dimensions.}
   \label{fig:NONCVEPerplecity}
%\vskip -10pt   
\end{figure*}

The training data $(X,y)$, where $X$ is a $\mathbb{R}^{133}$ vector consisted of $\mathbb{R}^{128}$ auto-encoded embedding, and the five `custom-made' features described above. We implemented upsampling with SMOTE at 20 percent level and used  random forests, GBM and two layer neural networks (all optimally calibrated.) 

%The performance of the random forest at 30 percent upsample level is displayed in Table \ref{tab:liftrf} below.
\begin{table}[ht]
\setlength{\tabcolsep}{3pt}
\begin{center}
%\captionsetup{justification=centering}
\caption{\textsc{Classifier performance across different embeddings}}\label{tab:liftcomp}
\vspace{3pt}
\ra{1.2}
\setlength{\tabcolsep}{6pt}
    \begin{tabular}{llll}
\toprule
\textbf{Embedding} &  $\mathbb{R}^{32}$    & $\mathbb{R}^{64}$      & $\mathbb{R}^{128}$\\ 
\midrule

%\cmidrule(lr){2-4}
%\cline{1-1} \cline{2-2} \cline{3-3} \cline{4-4} 
\textbf{GBM} &&&\\
\textbf{Lift} &        0.792  &   0.821  & 0.846\\
\textbf{AUC} &         0.938  &   0.950 &  0.957\\
\textbf{Accuracy}    &0.986   &  0.987  & 0.988\\
\textbf{Precision}   &0.244   &  0.269  & 0.297\\
\textbf{Recall}      &0.475   &  0.525  & 0.518\\
%\hdashline
\cmidrule(lr){1-4}
\textbf{Random Forest} &&&\\
%&32          64    &    128
\textbf{Lift}    &   0.720   &    0.728  &   0.726\\
\textbf{AUC}     &   0.909   &    0.918  &   0.915\\
\textbf{Acc}     &    0.954   &    0.962  &   0.962\\
\textbf{Prec}    &    0.086   &    0.105  &   0.101\\
\textbf{Recall}  &    0.574   &    0.587  &   0.613\\
\cmidrule(lr){1-4}
\textbf{Neural Network} &&&\\
%         &32       &    64    &  128 \\
\textbf{Lift}    &   0.779  &   0.825  & 0.846\\
\textbf{AUC}     &   0.942  &   0.956  & 0.962\\
\textbf{Acc}     &   0.980  &   0.984  & 0.975\\
\textbf{Prec}    &   0.184  &   0.232  & 0.176\\
\textbf{Recall}  &   0.534  &   0.560  & 0.733\\
\bottomrule
    \end{tabular}
\end{center}
    \vskip -10pt
\end{table}

Performance across the three model frameworks is summarized in the Table \ref{tab:liftcomp} and the lattice plot Figure \ref{fig:GBM_gain}.

A neural network model upsampled with SMOTE at 20 percent yields a slightly better AUC of 0.962. See Figure \ref{fig:NONCVEPerplecity} for more details. The lift, however, is much more pronounced at 0.846. Also a recall of 0.733 is a great improvement over competing models.

Note Fig. \ref{fig:NONCVEAccuracy} the misbehavior at a single epoch where performance drops over a single epoch. There are a few plausible explanations of this mystery. One plausible explanation is instability of the gradient, another is that the network is stuck in a faux minimum but able to subsequently recover. Since this paper is not about neural networks, and since we observe long-term stability (with respect to AUC, accuracy, recall and perplexity,) we didn't find it necessary to go into the  painful details.

\section{Conclusion}
This empirical analysis demonstrates that static code analysis with machine learning might indeed prove to be effective in identifying malicious or sloppy code that contains bugs that expose computers and the data to unnecessary costly risks.

This scoring method of the Android kernel demonstrates that most known vulnerable function lie in the top tier of the score distribution. This suggests that the procedure could be generalized and may be used to identify vulnerabilities and improve the code vetting process. 

To wrap ones head around this outcome, with approximately 1.1 million functions, 1 percent consists of 11,000 functions, of which 1,915 are known to be CVEs; that is, approximately 17.4\% of the top 1\% %percentile 
of scores are known to be bad. One would be amiss not to contemplate the consequences of having information about a cohort's risk of causing major disruption and how one should respond to such information.  
\section{Declaration of Competing Interests}
The authors declare that they have no known competing financial interests or personal relationships that could have appeared to influence the work in this paper.
\section{Credit authorship contribution}
\textbf{Joe Bar:} Conception, methodology, writing-original draft, resources and supervision.
\textbf{Peter Shaw:} Writing, methodology, review \& editing.
\textbf{Tyler Thatcher:} Experimentation and investigation.
\section{Acknowledgements} Special thanks to Acronis SCS (USA) for supporting this project, and our gratitude for the support and advice we've received from our friends including Sergey Ulasen and Sanjeev Solanki of Acronis (Singapore), Neil Proctor and Katerina Archangorodskaja of Acronis SCS (USA).  Peter Shaw was supported in part by the Jiangsu province, China, 100 Talent project fund (BX2020100).

%Bibliography
\bibliographystyle{unsrt}  
\bibliography{references} 

\section{Appendix}
%\vskip -25pt
%https://tex.stackexchange.com/questions/214618/author-biographies-in-elsarticle

\section*{Author biographies}
\setlength\parfillskip{0pt}\par\setlength\parfillskip{0pt plus 1fil}
%\vspace{-8pt}
%\begin{wrapfigure}[1]{l}{23mm}
%\end{wrapfigure}

\begin{wrapfigure}[10]{l}{23mm}
  \vspace{-8pt}
    %{\includegraphics[width=25mm,clip,keepaspectratio]{PSphoto.jpg}}
    {\includegraphics[width=25mm,clip,keepaspectratio]{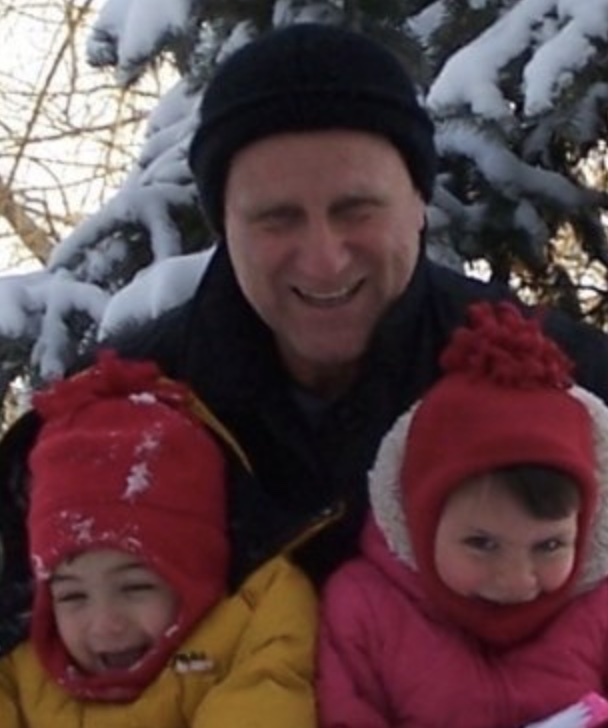}}
\end{wrapfigure}

\textbf{Joseph R. Barr} is Senior Director of Research at Acronis SCS (USA), (\url{www.acronisscs.com}) an industry leader in anti-ransomware and cybersecurity solutions. As head of research Joe's responsibility is to develop the methodology and the tools to help automate the process of identifying vulnerabilities in source code. Earlier Joe was Chief Analytics Officer at HomeUnion (\url{www.homeunion.com}) (sold to \url{www.Mynd.co}) where he was responsible for analytics and back-office quantitative analysis. Prior to that he was Chief Data Scientist at ID Analytics (now part of Lexis-Nexis \url{https://risk.lexisnexis.com/corporations-and-non-profits/credit-risk-assessment}) where he was responsible for the development of fraud-prevention and consumer credit risk products. Joe has begun his career as a mathematics professor at California Lutheran University. Joe has a Doctorate in mathematics from the University of New Mexico. He publishes extensively in the area of machine learning and combinatorics. 

%received his BA, Mathematics \& Statistics, University of Haifa, Haifa, Israel, and his PhD in Mathematics at the University of New Mexico, Albuquerque, NM (1991). He also has a MS, Mathematics, Florida State University, Tallahassee, FL (1985). Joseph Barr is a business leader with expertise in data science. Developed over a dozen commercial products with substantial market value. Engaging in scholarly research in machine learning.

%https://tex.stackexchange.com/questions/214618/author-biographies-in-elsarticle
%\begingroup
%\setlength{\intextsep}{0pt}%
%\setlength{\columnsep}{0pt}%

\vspace{1pt}
  \begin{wrapfigure}[10]{l}{23mm} 
  \vspace{-7pt}
    {\includegraphics[width=25mm,clip,keepaspectratio]{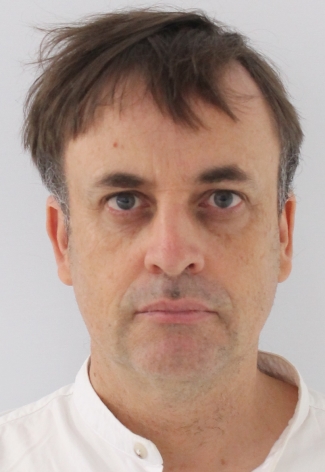}}
  \end{wrapfigure}

  \noindent\textbf{Peter Shaw} was born in Australia in 1969. He received the B.CompSci. Hon I and PhD degree in computer science from the Newcastle University Australia. He lectured in software engineering at Charles Darwin University between 2010 and 2018. He is currently a professor in the AI department at Nanjing University of Information Science and Technology (NUIST), China, and a Honoree Research Fellow at Menzies School of Child Health, Australia. His core research is fundamental advantages in AI using Fixed Parameter Tractable algorithms.

%\vspace{55pt}
\newpage
%\vspace{-27pt}
%\begin{minipage}
\setlength\parfillskip{0pt}\par\setlength\parfillskip{0pt plus 1fil}
\begin{wrapfigure}[10]{l}{23mm}
\vspace{-50pt}
\includegraphics[width=25mm,clip,keepaspectratio]{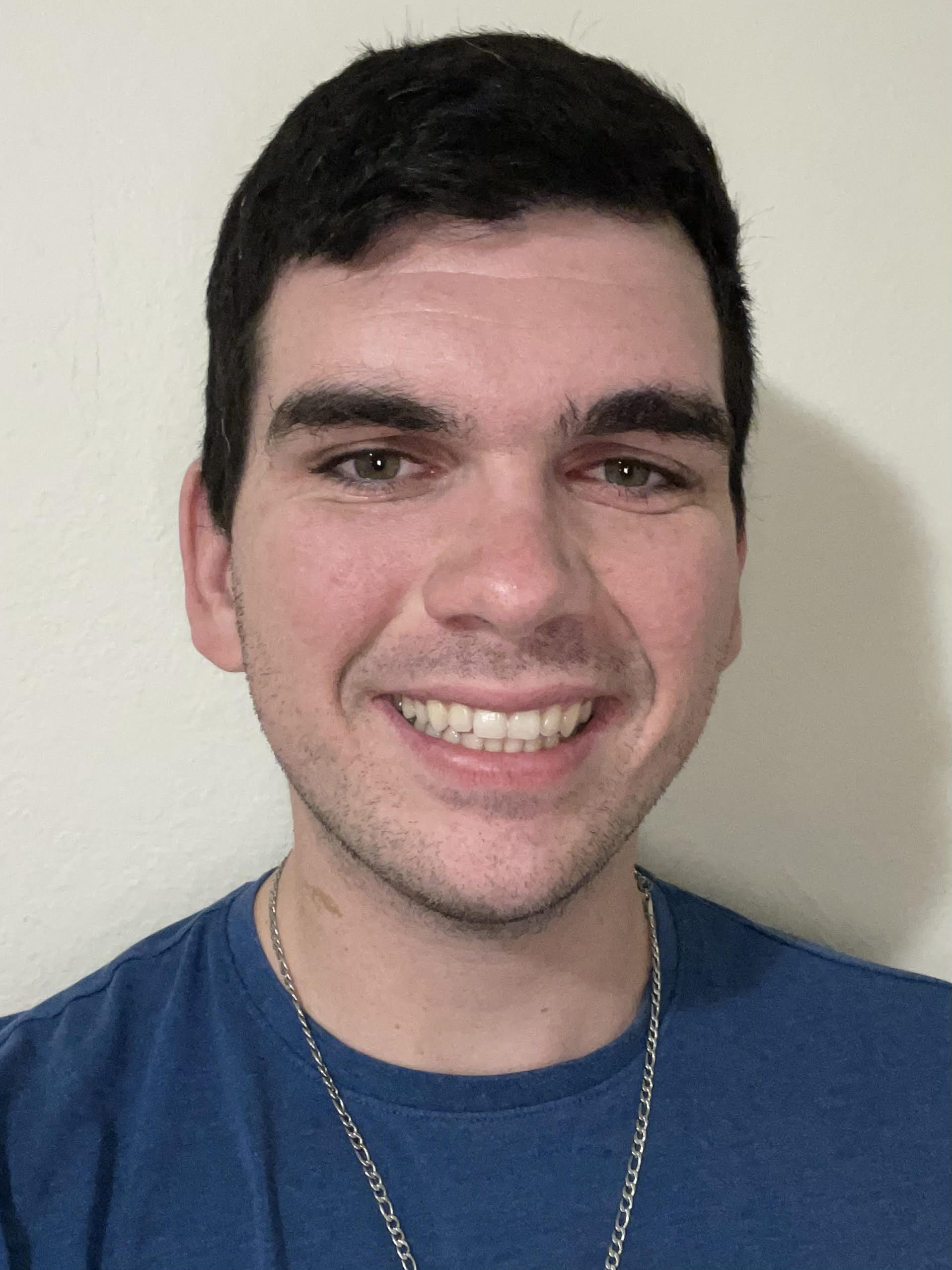}
\end{wrapfigure}

\noindent\textbf{Tyler Thatcher} is a Master’s student at NAU. He Currently works at Acronis SCS as a Machine Learning Engineer, researching and developing innovative ways to leverage machine learning in the cyber security domain. Previously, he  worked as a researcher for USGS Astrogeology and NAU SICCS.  
%\end{minipage}

\end{document}